\documentclass[conf]{new-aiaa}
\usepackage[utf8]{inputenc}
\usepackage{booktabs}

\usepackage{graphicx}
\usepackage{placeins}
\usepackage{amsmath}
\usepackage{subcaption}
\usepackage[normalem]{ulem}
\usepackage[version=4]{mhchem}
\usepackage{siunitx}
\usepackage{longtable,tabularx}
\usepackage[table]{xcolor}
\usepackage{tikz}
\usepackage{bbm}
\usepackage{pifont}

\newcommand{\ovalbox}[1]{\tikz[baseline=(char.base)]{
            \node[shape=rectangle,draw,inner sep=2pt] (char) {#1};}}

\title{Conditioning Aircraft Trajectory Prediction on Meteorological Data with a Physics-Informed Machine Learning Approach}

\author{Amy Hodgkin\footnote{Research Associate, Project Bluebird, The Alan Turing Institute} and Nick Pepper\footnote{Senior Research Associate and Co-Investigator, Project Bluebird, The Alan Turing Institute}}
\affil{The Alan Turing Institute, London NW1 2DB, United Kingdom}
\author{Marc Thomas\footnote{Co-Investigator, Project Bluebird, NATS}}
\affil{NATS, Whiteley, Fareham PO15 7FL, United Kingdom}

\begin{document}

\maketitle

\begin{abstract}
Accurate aircraft trajectory prediction (TP) in air traffic management systems is confounded by a number of epistemic uncertainties, dominated by uncertain meteorological conditions and operator specific procedures. Handling this uncertainty necessitates the use of probabilistic, machine learned models for generating trajectories. However, the trustworthiness of such models is limited if generated trajectories are not physically plausible. For this reason we propose a physics-informed approach in which aircraft thrust and airspeed are learned from data and are used to condition the existing Base of Aircraft Data (BADA) model, which is physics-based and enforces energy-based constraints on generated trajectories. A set of informative features are identified and used to condition a probabilistic model of aircraft thrust and airspeed, with the proposed scheme demonstrating a 20\% improvement in skilfulness across a set of six metrics, compared against a baseline probabilistic model that ignores contextual information such as meteorological conditions.

\end{abstract}

\section{Introduction}
\lettrine{T}rajectory prediction (TP) methods underpin many of the tools currently used in air traffic management (ATM) operations. However, a number of confounding factors limit the accuracy of TP such as uncertainties concerning aircraft mass, wind conditions, aircraft performance settings and pilot behaviours \cite{lymperopoulos2010sequential}. A natural way to handle uncertainties in such operational environments is to couple existing physics-based models with probabilistic methods that can propagate uncertainties. As an example, Matsuno and Matsuda \cite{MATSUNO2025104999} performed uncertainty quantification using ensemble forecast data to capture the effects of uncertain meteorological conditions. However, such an approach is limited by the availability of probability distributions for those model inputs understood to be uncertain, such as meteorological conditions. More importantly, the quality of predictions are potentially limited by misspecification within the physics-based TP model. What is required is a method to probabilistically generate trajectories over long horizon times, in order to capture the effects of epistemic uncertainties, while also ensuring that the generated trajectories are physically plausible.

Given sufficient training data, it is possible to treat TP as a sequence-to-sequence modelling task and use techniques such as long-short term memory networks (see, e.g. \cite{LSTM, LSTM2, shi20204}) and generative adversarial networks (see, e.g. \cite{WU2022103554, pang2020conditional}) to perform short-term predictions based on observations of a trajectory. Alternatively, models can be applied auto-regressively, for example Barratt et al. used Gaussian mixture models to generate trajectories in the terminal manoeuvring area of John F Kennedy International Airport \cite{barratt2018learning}. Such techniques are effective in highly procedural environments, an example being terminal control, but are of limited utility in operational settings where aircraft behaviours are driven by the clearances issued by air traffic control officers (ATCOs), such as en route airspace. En route airspace may also require predictions over longer time horizons, incorporating manoeuvrers that may have not yet been initiated at the time of prediction, for which black box machine learning (ML) methods are ill suited. 

\begin{figure}[h]
    \centering    \includegraphics[width=0.5\linewidth]{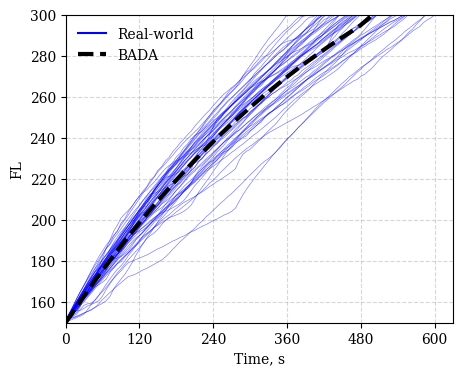}
    \caption{BADA prediction of B738 climbs, compared to 50 trajectories observed in real-world operations.}
    \label{fig:badaexample}
\end{figure}

Architectures in which components of an existing physics-based model are machine learned from historic data are  more suitable for en route airspace {because the physics-based model can explicitly model ATCO clearances and procedures}. In the proposed framework, machine learned terms are passed through a physics-based model, such as BADA, ensuring that generated trajectories satisfy physical constraints, for example enforcing monotonic altitude increases during climb. Such techniques are now commonplace in other fields, for instance in machine learning  turbulence closure models (see, e.g. Parish and Duraisamy \cite{parish2016paradigm}). Recently, similar methods have been proposed for the probabilistic generation of aircraft trajectories, with Pepper and Thomas \cite{pepper2023learning} introducing a method for generating probabilistic aircraft climbs, which was later extended to incorporate descents by Hodgkin et al. \cite{hodgkin2025probabilistic} and full 4D predictions in en route airspace by Pepper et al. \cite{pepper4984556probabilistic}. Although regulatory guidance for incorporating machine learning models into operational systems is still evolving (see, e.g. \cite{faa_roadmap, easa_guidance, caa_guidance}), it is clear that the safety-critical nature of TP in ATM means that {the range of possible outputs and accuracy of any machine learned model must  be understood. An advantage of the proposed physics-informed machine learning (PIML) model is that it is more closely aligned with existing frameworks for validation and verification than TP models employing black-box machine learning algorithms.

Despite the progress that has been made with incorporating data-driven terms in physics-based TP models, a limitation of previous work is that trajectory samples are generated without being conditioned on factors such as meteorological conditions and aircraft operators. These factors are understood to strongly influence aircraft performance. The work presented here has been designed to overcome these limitations, and is intended for trajectory generation and prediction within an airspace digital twin \cite{DTpaper}. This paper makes several contributions to the literature: 

\begin{itemize}
    \item \textbf{Identification} of an informative set of features that can be used to condition data-driven TP models.
    \item \textbf{Improved accuracy} of mean predictions from a PIML TP model, by conditioning trajectory generation on these informative features.
    \item More \textbf{skilful generation} of trajectories by conditioning probabilistic trajectory generation on informative features.
\end{itemize}

The remainder of this paper is structured as follows: Section~\ref{sec:methods} outlines the proposed PIML method. Section~\ref{sec:data} describes a dataset of historic aircraft trajectories and forecast meteorological data associated with these trajectories that was used for model training and testing. Section~\ref{sec:fi_proj} describes the feature importance (FI) study to identify a set of informative features based on contextual information. Finally, Section~\ref{sec:results} investigates two options for conditioning probabilistic trajectory generation on these informative features within the proposed framework. 

\section{Methods}\label{sec:methods}
Figure~\ref{fig:schematic} is a schematic illustrating the proposed method for generating trajectories conditioned on weather, operator and other relevant information. There are several components to the model. Firstly, a set of informative features is used as an input vector to a probabilistic ML model which has been trained on historic flight data. Samples from this model are drawn in a low-dimensional latent space. Each dimension of this space is associated with a weight in an orthonormal basis for aircraft thrust and calibrated airspeed (CAS). Therefore, samples from the probabilistic models can be mapped to a thrust and CAS function that in turn condition the Base of Aircraft Data (BADA) model, a physics-based model that generates a complete aircraft trajectory. This section describes each component in greater detail. 

\begin{figure}
    \centering
    \includegraphics[width=0.9\linewidth]{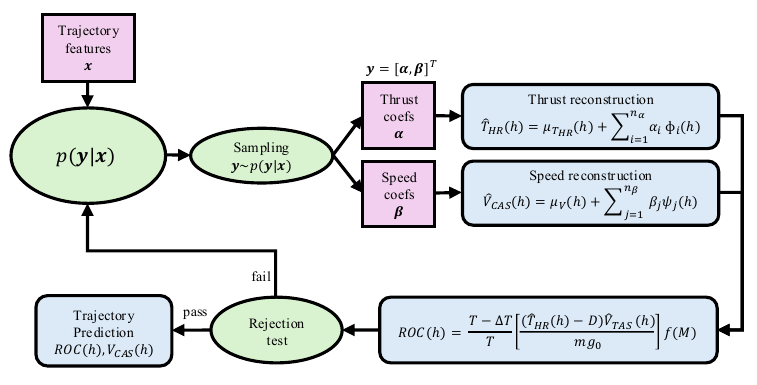}
    \caption{Schematic of the proposed PIML model.}
    \label{fig:schematic}
\end{figure}

\subsection{The BADA model}\label{sec:bada}

The BADA model provides a set of physics-based equations and aircraft specific parameters for trajectory prediction. It is the physics-based component of the proposed scheme. The total-energy equation, which relates the change in kinetic and potential energy to the work done on the aircraft, is used to define a function for rate of climb/descent (ROCD) \cite{nuic2010user}:

\begin{equation}
    \frac{dh}{dt} = \frac{T-\Delta T}{T} \Big[ \frac{(T_{HR}-D)V_{TAS}}{mg_0} \Big] f(M), \label{eq:bada_rocd}
\end{equation}
where $\frac{dh}{dt}$ denotes the ROCD, $T_{HR}$ is the aircraft thrust parallel to the velocity vector, $D$ is the drag, $V_{TAS}$ is the aircraft true airspeed (TAS), $m$ is the aircraft mass, $\frac{T-\Delta T}{T}$ represents a temperature correction factor that may be applied to BADA, although in what follows this is set to 1. $f(\cdot)$ is the energy share factor which determines energy availability in the speed regime the aircraft is operating in as a function of Mach number ($M$). The term \emph{speed regime} refers to whether the aircraft is climbing or descending at constant CAS or Mach and whether it is above or below the geodetic altitude of the tropopause. Several of the terms within \eqref{eq:bada_rocd} are themselves functions of TAS and $h$, for instance BADA provides equations for $T_{HR}$ as a function of $h$ for jet aircraft, while $D$ is a function of geodetic altitude and $V_{TAS}$, whose form is dependent on the mode of operation of the aircraft. The BADA model uses the international standard atmosphere (ISA) model to convert quantities such as CAS, which are independent of altitude, to TAS at a given altitude. 

BADA is a deterministic model, calibrated using} a table of parameters, such as aircraft mass, that are specific to an aircraft type and are available under licence. While the parameters for each aircraft type are calibrated to reproduce nominal performance characteristics based on reference data, the BADA model alone cannot capture the variability in climb and descent performance in real-world operations. This is demonstrated in Figure~\ref{fig:badaexample}, which shows the nominal BADA trajectory prediction against a set of 50 real-world trajectories. The standard distribution of the time for the trajectories to reach flight level (FL) 300 is 55.5 seconds, a considerable amount of uncertainty given the mean length of the BADA trajectory is 499 seconds. To capture the uncertainty seen in real-world trajectories, a functional representation of the misspecification between predicted and observed trajectories is used as a reduced-order representation of a trajectory, with a probabilistic, machine learned model trained to generate trajectories in this space. The next sub-section describes this method. 

\subsection{Reduced-order representation of thrust and speed}\label{sec:low_order}
In the proposed PIML scheme, aircraft thrust and CAS is learned as a function of $h$. This allows a machine learning model to apply a probabilistic, data-driven enhancement to BADA. To reduce the dimensionality of the learning task these continuous functions, denoted $\hat{T}_{HR}(h)$ and $\hat{V}_{CAS}(h)$, are projected onto a set of orthonormal basis functions:
\begin{subequations} \label{eq:fpca_sum}
\begin{equation}
    \hat{T}_{HR}(h) = \mu_{T_{HR}}(h) + \sum_{i=1}^{n_\alpha} {\alpha}_{i} \phi_i (h) , \label{eq:fpca_drag}
\end{equation}
\begin{equation}
    \hat{V}_{CAS}(h) = \mu_V(h) + \sum_{j=1}^{n_\beta}{\beta}_{j} \psi_j (h), \label{eq:fpca_cas}
\end{equation}
\end{subequations}
where $\mu_{T_{HR}}$ and $\mu_V$ are mean functions and $\boldsymbol{\alpha}\in\Re^{n_\alpha}$ and $\boldsymbol{\beta}\in\Re^{n_\beta}$ are a set of model weights. The basis functions are orthonormal and therefore satisfy the conditions: 
\begin{equation}
    \int \phi_i(h) \phi_j(h) dh = \delta_{ij}\; \text{and} \;\int \psi_k(h) \psi_l(h) dh = \delta_{kl},
\end{equation}
with $i,j\in[1,n_\alpha]$ and $k,l\in[1,n_\beta]$. The basis functions are identified through functional principal component analysis (fPCA) of a training dataset of thrust and CAS data \cite{ramsay2013functional}. The basis functions may be continuous, user defined functions, such as $\beta$ splines, or they can be discrete functions. Given the abundance of training data in the dataset, we take the more flexible approach of defining the basis functions as discrete functions over a grid of altitudes, $\boldsymbol{g}\in\Re^{n_g}$. 

The summations in \eqref{eq:fpca_sum} are truncated. The number of terms of the summation is determined by the minimum number of terms needed to achieve 80\% cumulative explained variance. A unique set of basis functions is found for each aircraft type as the service ceiling can vary significantly between aircraft types. In this work a further processing step is taken to split trajectories of climbing aircraft into `sub-trajectories'. Figure~\ref{fig:notation} in Appendix~\ref{app:fpca} illustrates this concept. This step is necessary because aircraft are frequently `step climbed' when airspace is busy. Meteorological conditions can therefore vary significantly between individual climbs within the same trajectory. The use of sub-trajectories requires a sequential-least squares algorithm to project onto the orthonormal basis, described in Appendix~\ref{app:fpca}. This appendix also provides more details on the required data processing, which yields the set of fPCA weights $\mathcal{D}_{y} = [\boldsymbol{y}_1,\dots, \boldsymbol{y}_{n_s}]$, where $\boldsymbol{y}_i=[\boldsymbol{\alpha}^*_i, \boldsymbol{\beta}^*_i]^\top$. These are the weights associated with a set of sub-trajectories $\mathcal{D}_{s} = [\mathcal{S}_1, \dots, \mathcal{S}_{n_s}]$. 

\subsection{Probabilistic Machine Learning}\label{sec:prob_model}
The proposed method described in Figure~\ref{fig:schematic} maps contextual information in a vector of features, denoted $\boldsymbol{x}\in\Re^{n_x}$, to a distribution of fPCA weights, $p(\boldsymbol{y}|\boldsymbol{x})$, from which probabilistic trajectories may be generated. This paper evaluates the efficacy of two popular probabilistic machine learning methods for this task. This section provides a high-level overview of these methods, with specific implementation details described further in Section~\ref{sec:results}. 

\hfill \\
\noindent \textbf{Gaussian Processes} \\
A Gaussian process (GP) is a non-parametric, Bayesian modelling approach that can flexibly model input-output mappings while providing well-calibrated estimates of predictive uncertainty. Although there are multi-output implementations of GPs, in this work we simplify the modelling task by exploiting the formulation of $\boldsymbol{y}$ to model each component independently. This approximation is justified because projection onto the fPCA basis decorrelates the outputs. A set of GPs are trained to learn the mapping from $\boldsymbol{x}\rightarrow\boldsymbol{y}_i,\;i\in[1,n_\alpha+n_\beta]$:
\begin{equation}
    p(\boldsymbol{y}_i|\boldsymbol{x}) \sim N(\mu^{(i)}(\boldsymbol{x}), \sigma^{(i)}(\boldsymbol{x}, \boldsymbol{x}')),
\end{equation}
with:
\begin{align}
    \mu^{(i)}(\boldsymbol{x}) &= \mu_0^{(i)}(\boldsymbol{x})+K(\boldsymbol{x}', X)^\top K(X,X)^{-1}\big(Y^{(i)}-\mu_0^{(i)}(\boldsymbol{x})\big),\\
    {{\sigma^2}^{(i)}}(\boldsymbol{x}) &= K(\boldsymbol{x}',\boldsymbol{x}') -K(\boldsymbol{x}',X)^\top K(X,X)^{-1}K(\boldsymbol{x}',X),
\end{align}
where $X\in\Re^{n_x\times n_s}$ collects the feature vectors associated with each sub-trajectory, $\mu_0^{(i)}$ is a user specified mean function (for the i\textsuperscript{th} element of $\boldsymbol{y}$) and $Y^{(i)}\in\Re^{n_s}$ collects the $i\textsuperscript{th}$ element of each target vector in $\mathcal{D}_y$. The $p\textsuperscript{th}$ and $q\textsuperscript{th}$ elements of the covariance matrix, $K(X,X)$, are determined through:
\begin{align}
    K_{pq}(X,X) = k\big(\boldsymbol{x}^{(p)}, \boldsymbol{x}^{(q)}\big)+\epsilon_i \delta_{pq}, \; p=[1,n_s],\,q=[1,n_s],
\end{align}
where $k(\cdot)$ is a covariance function specific to each GP, $\epsilon_i$ a constant to ensure numerical stability in the inversion of $K$, and $\delta$ is the Kronecker delta. The functional form of the covariance function is determined by the user. In this work, best results were achieved using a Matérn 5/2 kernel. Lengthscales for each input dimension were tuned to fit the training data using the predictive log likelihood \cite{jankowiak2020parametric}.

A well documented limitation of GPs is that inference depends on the inversion of a covariance matrix with dimensions $n_s \times n_s$ (the number of samples in the training data). In this paper, the inducing point framework of Hensman et al. is used to mitigate this cost and allow GPs to efficiently learn from large datasets of trajectory data \cite{hensman}. The interested reader is referred to the canonical reference, Rasmussen and Williams, for more details on GPs \cite{GP1}. The GP was implemented using the \texttt{GPytorch} Python package \cite{GPytorch}, with any categorical features K-fold target encoded\footnote{Best results were achieved with $K=3$ as this ensured that there was sufficient data within each fold for the less frequently occurring aircraft types.} \cite{pargent2019benchmark} and up to 1,000 inducing points used for each GP.

\hfill \\
\noindent \textbf{Deep Ensembles} \\
Deep ensembles (DEs) may be thought of as an analogue to the random forest for neural networks, using an ensemble of neural networks to achieve improved generalisation performance through the ensembling of predictions from decorrelated models \cite{NIPS2017_de}. A set of $n_e$ multi-layer perceptrons (MLPs) are trained with varied initial weights and potential architectures to achieve randomness. The output of each network is a set of independent Gaussians for each element of $\boldsymbol{y}$, which are ensembled through:
\begin{align}
    &\mu^{(i)} (\boldsymbol{x}) = \sum_{k=1}^{n_e}\mu_{i,k}(\boldsymbol{x}),\\
    &\sigma^{(i)}(\boldsymbol{x})= \sum^{n_e}_{k=1}\frac{1}{n_e}\mu^2_{i,k}(\boldsymbol{x})-\mu^{(i)}(\boldsymbol{x})+\frac{1}{n_e}\sum_{k=1}^{n_e}\sigma_{i,k}^2(\boldsymbol{x}),
    \label{eq:de_var}
\end{align}
where $\mu_{i,k}$ and $\sigma_{i,k}$ represent the outputs of the $k\textsuperscript{th}$ ensemble member for the $i$\textsuperscript{th} element in $\boldsymbol{y}$. The first summation in \eqref{eq:de_var} represents uncertainty arising from discrepancy between the ensemble members, while the second summation captures the contribution to the uncertainty from each individual ensemble member. Each ensemble member is trained to minimise the negative log likelihood loss over the training data. The DEs were implemented in \texttt{tensorflow} using the code of Ho et al. \cite{HO2024106443}. A sweep over $n_e$ and the architecture of the MLPs was performed for the B738 and BE20 aircraft types, with best results achieved using $n_e=8$ and three hidden layers with a [128, 256, 128] pattern of neurons. 

\hfill \\
\noindent \textbf{Gaussian fitting} \\
The models presented here were benchmarked against the probabilistic model proposed in Hodgkin et al. \cite{hodgkin2025probabilistic}, in which expectation maximisation (EM) was used to fit a multivariate normal distribution to the fPCA subspace for available training data. This provided a useful baseline against which to evaluate the skilfulness of the GP and DE approaches as samples of $\boldsymbol{y}$ are generated independently of the features.

\subsection{Rejection Test}
The investigated ML models sample from continuous probability distributions with infinite support. Occasionally, samples are drawn which generate a trajectory with a ROC of less than 500 ft/min for at least one radar blip. To ensure all generated trajectories achieve the minimum legal climb rate, these trajectories are removed from the sample set. The rejection rate is the percentage of trajectories removed. The lower the rejection rate the more efficient the ML model will be at generating a plausible sample.

\section{Data Preparation}\label{sec:data}

The dataset consists of two components: a dataset of trajectories harvested from Mode S radar surveillance returns and weather forecasts from the UK Met office dataset. Data spanning 60 days, from May 2019 to April 2020, was selected for use. Five days are selected at random from each month to minimise bias arising from seasonal effects. 

\subsection{Trajectory Data}
This paper aims to exclude terminal airspace procedures, and has therefore focused analysis on flight levels above FL150, which generally corresponds to en-route operations in UK airspace. This paper focusses on ten aircraft types that feature a range of performance characteristics and vary in how frequently they occur within UK airspace. Following the developments of Appendix~\ref{app:fpca}, trajectories were processed into sub-trajectories to isolate climbs that are continuous portions of a trajectory with $\frac{dh}{dt}\geq 500 \text{ft/min}$. Table \ref{tab:data_amount} details the investigated aircraft types and the numbers of both trajectories and sub-trajectories in the dataset. Employing the optimization process defined in \eqref{eq:opt1}, an optimal set of weights were found for each sub-trajectory in the dataset. Note that the service ceiling varies significantly by aircraft type, requiring a unique set of basis functions for each aircraft type. 

In addition to positional data and the ROCD, the Mode S data also contains addition information that is pertinent to TP. Table~\ref{tab:features} describes these additional features and explains their reason for inclusion in the FI study. Some features such as \emph{operator}, \emph{origin}, \emph{intent\_code}, and \emph{flight\_type} are categorical and were one-hot or target encoded, depending on what was most suitable for the choice of machine learning model. Time-related features such as \emph{time\_of\_day} were encoded in sin and cos pairings to retain their cyclical nature. 

Some categorical features can have high cardinality, potentially introduction noise into the FI study. To address this, an `other' label was assigned for infrequently occurring categories in the \emph{operator}, \emph{origin}, and \emph{intent\_code} features. The \emph{intent\_code} feature refers to UK Intention Codes, which provide a coarse description of the flows within UK airspace \cite{intent_code}. When coupled with the \emph{origin} feature, this provides a means of encoding the effect of local procedures that are likely to be followed by an aircraft. Both features are of less cardinality than the aircraft filed route, which would be an alternative method of encoding this information. Based on their destination airports, flights were categorised as domestic (within UK), superdomestic (within Europe) or long haul.
\begin{table}[]
    \centering
    \caption{Investigated aircraft types, the frequency with which they occur within the dataset, and the cardinality of some of the categorical features associated with that aircraft type.}
    \begin{tabular}{ccccccccc}
    \toprule
    \toprule
        Aircraft & No. of & No. of & No. of & No. origin & $n_\alpha$ & $n_\beta$ & Engine\\
        Type & Trajectories & Sub-trajectories & operators & airports &  &  & Type \\
        \midrule
        B738 &  30,182 &  36,594  &  4 &  7 &  12 &2   & Jet\\
        A320 &  21,022 &  26,057  &  5 &  5  & 16& 2&  Jet\\
        A319& 18,150 & 21,697 & 4& 7& 15 & 2 & Jet\\
        E190 & 4934& 5980 & 5& 4& 5& 2  & Jet \\
        DH8D &  1935 &  2131 &  3 &  9  & 5&2  & Turboprop\\
        A388 & 1552& 2094 & 6& 5& 22& 3  & Jet\\
        E170 & 775& 897 & 3& 4& 14 & 2  & Jet\\
        PC12 & 360& 461& 3& 5& 6&3 & Turboprop\\
        C56X & 330& 371 & 4& 6& 9&4  & Jet\\
        BE20 &  318 &  331 & 4& 5& 3& 2 & Turboprop\\
        \bottomrule
        \bottomrule
    \end{tabular}
    \label{tab:data_amount}
\end{table}

\begin{table}[]
    \centering
    \caption{Descriptions of the investigated features. (*features which are encoded as statistics)}
    \begin{tabular}{cp{4cm}p{4cm}c>{\centering\arraybackslash}p{2cm}}
        \toprule
        \toprule
        Feature & Description & Reason for inclusion & Encoding & Retained? \\
        \midrule
        \emph{operator} & Airline operating aircraft & Operator procedures differ. Cruise speed is specified in filed flight plan.  & One-hot/Target & \checkmark \\
        \emph{origin} & Origin airport & Local procedures may influence trajectory & One-hot/Target & \checkmark \\
        \emph{intent\_code} & Point at which aircraft will leave the London FIR, combines multiple routes \cite{intent_code}  & Followed route may impact procedures. Can be cross-referenced with \emph{wind\_along\_track} & One-hot/Target  & \checkmark \\
        \emph{flight\_type} & Flight type categorised as domestic, super-domestic (within Europe) or long-haul & An indicator of flown distance and potentially aircraft mass & One-hot/Target & \checkmark\\
        \emph{month\_of\_year} & Month of year & Potential seasonal influence on procedures & Sin/cos  & \checkmark\\
        \emph{day\_of\_week} & Day of the week & Baseline feature & Sin/cos & $\times$\\
        \emph{time\_of\_day} & Time of day, rounded to the hour & Potential influence of time-dependent procedures & Sin/cos & $\times$\\
        \emph{fl\_min} & Minimum FL & ROCD and CAS are functions of altitude in physics-based TP models & - & \checkmark  \\
        \emph{fl\_max} & Maximum FL & ROCD and CAS are functions of altitude in physics-based TP models & - & \checkmark  \\
        \emph{fl\_range} & Range of FLs through which aircraft climbs & ROCD and CAS are functions of altitude in physics-based TP models & - & \checkmark  \\
        \emph{wind\_magnitude}* & Nearest-neighbour wind forecast & Absolute wind strength may impact selected CAS & - & \emph{time\_grad} \\
        \emph{wind\_along\_track}* & Wind vector along the aircraft track & Head and tail wind impact fuel efficiency which may impact selected CAS & - & \emph{mean} \& \emph{time\_grad}\\
        \emph{wind\_cross\_track}* & Wind magnitude across the aircraft track & Not expected to influence, covered by wind\_magnitude& - & $\times$\\
        \emph{temp}* &  Difference in forecast temperature from ISA temperature across aircraft track & Important parameter in ISA model used by BADA & - & \emph{mean}, \emph{std} \& \emph{time\_grad}\\
        \bottomrule
        \bottomrule
    \end{tabular}

    \label{tab:features}
\end{table}

\subsection{Forecast data}
\begin{figure}
    \centering
\includegraphics[width=0.6\linewidth]{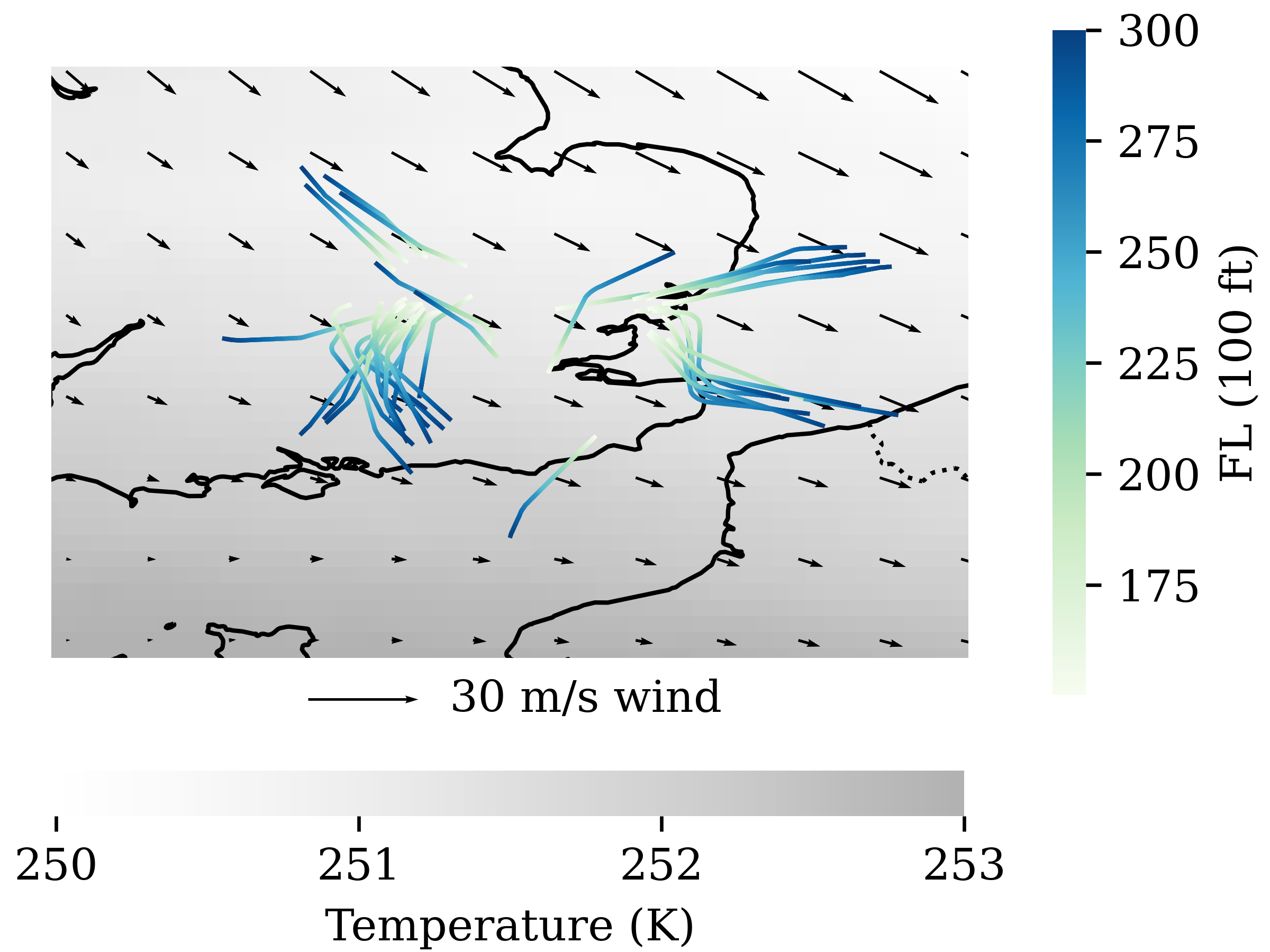}
    \caption{Mean forecast wind and temperature field, overlaid with trajectories of climbing aircraft originating at London airports for a 3 hour period from 07:30UTC to 10:30UTC on 3 July 2019 between FL150 and FL300.}
    \label{fig:met_traj_map}
\end{figure}

Forecast data spanning 60 days from the UK Met Office were used for the study, with a spatial resolution of {0.11 degrees in latitude and longitude, 100 hPa of isobaric pressure (converted to altitude in m)} and with three hour intervals. Zero-order interpolation was used to associate each radar blip in the trajectory data with the northwards ($u$) and eastwards ($v$) component of the forecasted wind field, as well as the temperature, $T$. By using the aircraft heading, which is reported in Mode S radar returns, the along and cross-track components of the forecast wind field were computed for each radar blip. Figure~\ref{fig:met_traj_map} shows a slice of a Met office forecast overlaid with trajectory data. The wind and temperature values plotted are the mean forecast for a three hour window from 07:30UTC to 10:30UTC on 3 July 2019 between FL150 and FL300 and the trajectories plotted are from within these time and altitude limits.

\FloatBarrier
\section{Feature importance study}\label{sec:fi_proj}
A FI study was performed to determine a set of informative features which could be used to condition a probabilistic TP model. Random forest (RF) models were trained to perform the FI study due to their robustness to mixed feature types and stable importance estimates with minimal tuning. Three heuristic metrics were used for quantifying feature importance:
\begin{enumerate}
    \item Gini feature importance: a measure of feature importance specific to tree-models such as random forests. It is based on how frequently a feature is used to reduce impurity (uncertainty) in the trees of the random forest model.
    \item Feature permutation: permute the values for one feature at a time in the validation model. Importance is quantified by the decrease in $R^2$ compared to the RF model prediction without permutation. 
    \item Feature dropping: drop one feature at a time from the training data and retrain the RF model. Compare the decrease in $R^2$ between the new RF model and original to quantify feature importance. 
\end{enumerate}
An 80:20 training:test split was performed on the dataset for the FI study. The heuristic FI metrics involved analysing the coefficient of determination, $R^2$, for the held-out test set. To simplify the analysis the first coefficient of the fPCA representation was used. This was expected to be sufficient given that the first component explains the largest portion of the variance (see Figure~\ref{fig:expvar} in Appendix~\ref{app:b738_coeffs}). Further simplification was introduced by only studying the B738, which was the most commonly occurring aircraft type in the dataset. 

There are a variety of ways in which the wind and temperature field data may be encoded within a feature vector. Here we represent this data by computing statistics on the forecast conditions encountered by each aircraft during each sub-trajectory step. Here these features are computed through post-processing trajectory data. In a live system an initial coarse TP would be needed to generate a ground track and altitudes with which to compute these features.

\begin{figure}
    \centering
    \includegraphics[width=1\linewidth]{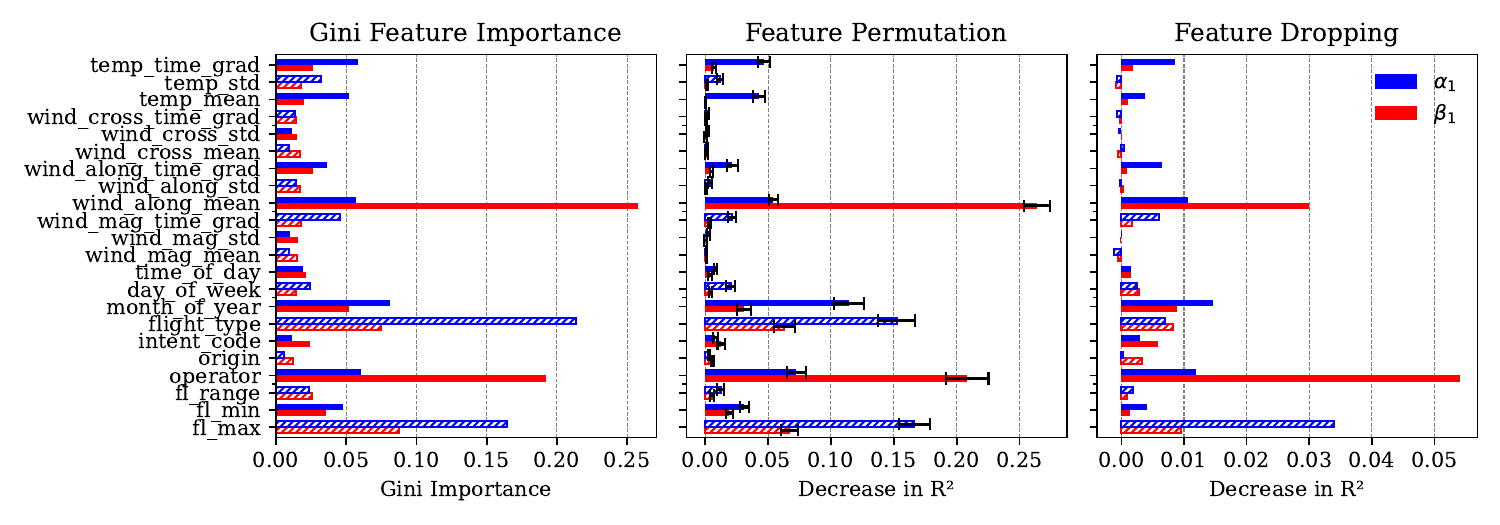}
        \caption{Feature importances for predictions of the first coefficient in the reduced-order representation of sub-trajectories for the B738.}
    \label{fig:fi_B738}
\end{figure}

The feature importance results for the B738 are shown in Figure~\ref{fig:fi_B738}. Blue bars indicate the FI for predicting $\boldsymbol{\alpha}_1$, which is associated with the generated aircraft thrust, while red bars represent the FI of $\boldsymbol{\beta}_1$, associated with the CAS. Features were permuted 10 times, with two standard deviations of the decrease in $R^2$ indicated by the error bars. Longer bars in Figure~\ref{fig:fi_B738} can be interpreted as the corresponding feature having greater importance. From the figure it is clear that along track wind and the operator type are by far the most important features for predicting $\beta_1$. \emph{fl\_max} has importance for predicting $\beta_1$ according to the Gini FI and permutation heuristics. This is also the dominant feature for predicting $\alpha_1$, which is understandable given that the \emph{fl\_max} will indicate whether the aircraft has passed the CAS-Mach transition point, beyond which aircraft typically climb at a slower rate. The most important features for predicting $\alpha_1$ are more associated with destination and route, rather than meteorological conditions. This is perhaps to be expected given that these features concerning the operational context will influence the fuelling and hence the mass of the aircraft, which is understood to have a significant influence on climb performance. 

Based on the results presented in Figure~\ref{fig:fi_B738}, a set of 14 features were retained, which are listed in Table~\ref{tab:features}.

\section{Results}\label{sec:results}
Having determined the set of informative features to use to learn the mapping $\boldsymbol{x}\rightarrow p(\boldsymbol{y}|\boldsymbol{x})$, probabilistic models were trained on the trajectory data. Recognising that the best choice of probabilistic model might vary between aircraft types, the relative skilfulness of the GP and DE relative to the Gaussian baseline model were computed for a set of metrics, using the held-out test dataset. 

\subsection{Metrics}
The time taken for the $i$\textsuperscript{th} sub-trajectory\footnote{{To facilitate the fair comparison of the probabilistic models, the projection of sub-trajectories in the low-order representation were used as a target, rather than the raw trajectory data, to minimise biasing effects from mapping between the space of fPCA weights and trajectories.}} in the test dataset, $\mathcal{S}_i$, to climb between $h^{(0)}$ and $h^{(n)}$ is denoted $\tau_i$. Metrics were computed using 500 samples of $\boldsymbol{y}$ drawn from the probabilistic models. Using \eqref{eq:fpca_drag} and \eqref{eq:fpca_cas} these samples were converted into sampled thrust and CAS functions that were used to condition BADA and generate 500 trajectories. We denote $F(t)$ as the cumulative distribution function (CDF) of the time taken for aircraft to reach the top of climb for these generated trajectories. Given the observation $\tau_i$ and $F(t)$ the cumulative ranked probability score (CRPS) can be computed for each probabilistic trajectory through: 
\begin{align}
    CRPS(\tau, F(t)) = \int_{-\infty}^\infty \big(F(t)- \mathbbm{1}(t-\tau)\big)^2dt,
\end{align}
where $\mathbbm{1}(\cdot)$ is the Heaviside step function. A CRPS will be associated with each trajectory in the test dataset and for each of the three probabilistic models described in Section~\ref{sec:prob_model}. These scores are aggregated into a skilfulness score, which is computed through:
\begin{equation}
    S=1-\frac{\sum_{k=1}^{n_t}CRPS_{GP}\big(\tau_k, F_{GP}(t)\big)}{\sum_{k=1}^{n_t}CRPS_{base}\big(\tau_k, F_{base}(t)\big)},
    \label{eq:skill}
\end{equation}
for the GP, with a similar expression for the DE, with $n_t$ representing the number of sub-trajectories in the test dataset. The sub-script `base' refers to the fitted Gaussian baseline model. Positive values of $S$ indicate that the probabilistic forecast of $\tau$ from the GP is more accurate than than that of the baseline. In other words, that conditioning the data-driven trajectory generator on contextual information yields a more realistic set of trajectories than the baseline, fitted Gaussian, method. 

The CRPS-based skilfulness score was computed for three quantities: CAS, rate of climb (ROC), and time to climb. Furthermore, \eqref{eq:skill} could be reformulated using the root mean square error (RMSE) instead of CRPS, providing complementary set of metrics focussing on the accuracy of the mean predictions. In contrast, the CRPS-based skilfulness score evaluates how well calibrated the probabilistic generation of trajectories is. There are therefore six metrics over which model performance can be measured.   

\subsection{Application to test dataset}
\begin{figure}
    \centering
    \includegraphics[width=1\linewidth]{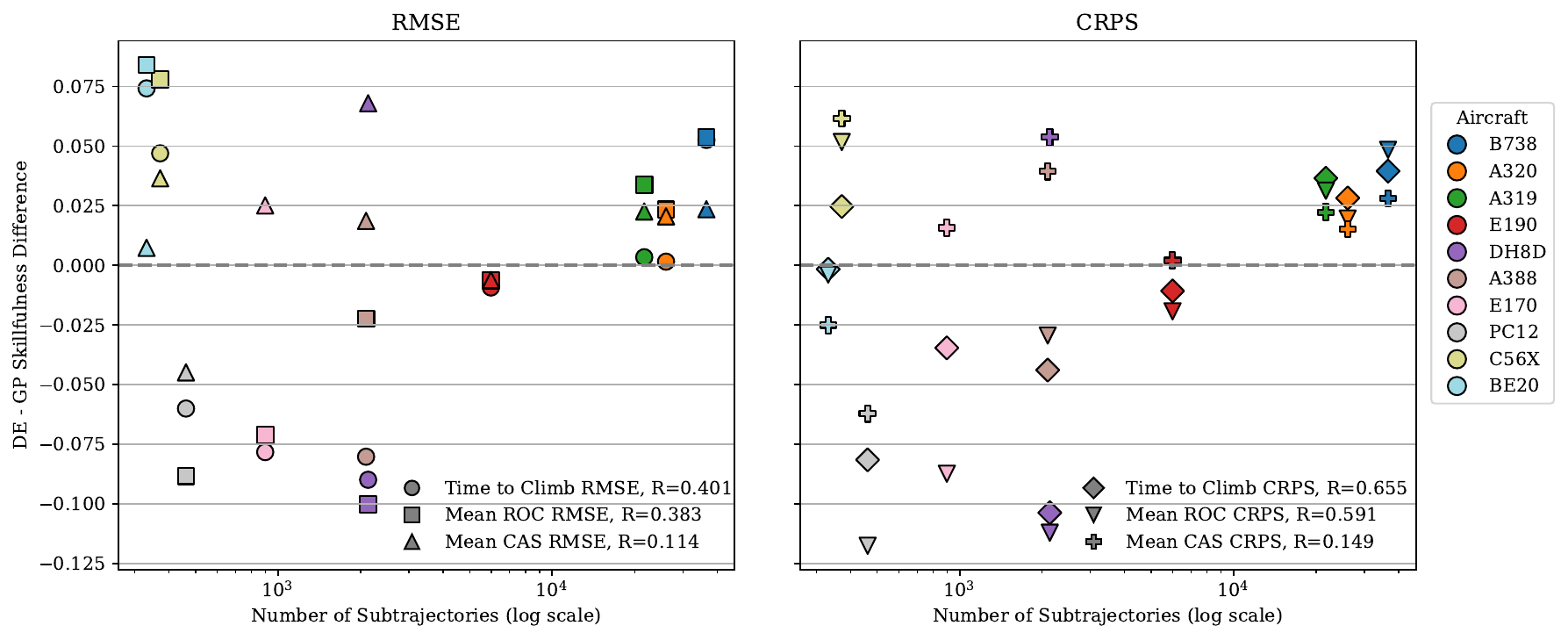}
    \caption{Difference in skilfulness score between the Deep Ensemble and Gaussian Process model. Points above the grey dashed line indicate a higher score for the Deep Ensemble. The correlation ($R$) is given in the legend for each metric.}
    \label{fig:score_diff}
\end{figure}

Figure~\ref{fig:score_diff} plots the difference in the skilfulness scores of the DE and GP for each of the metrics against the number of sub-trajectories in the training dataset. The left panel plots skill scores computed using the RMSE, while the right panel plots the skill scores computed using the CRPS. In other words, the left panel displays the relative skilfulness of the mean predictions of models and the right panel displays the skill score associated with probabilistic generation of trajectories. For those aircraft types with larger training datasets (around 20,000 sub-trajectories or greater), such as the B738, the DE tended to be more skilful than the GP, with values lying in the positive half plane of Figure~\ref{fig:score_diff}. On the other hand, less frequently occurring aircraft were more often modelled more skilfully by the GP, with skill scores lying in the negative half plane. This is unsurprising given the depth of the DE models and the relative scarcity of some of the less frequently occurring aircraft types.

Table~\ref{tab:skilfulness} displays the skilfulness for the GP and DE for each aircraft type and for each of the six metrics. For each aircraft type, the most skilful model for each metric is indicated by a box and red cells indicate those probabilistic models with negative skilfulness, i.e. those for which the baseline model is more accurate. The DE and GP generally have positive skilfulness values. 

The last row of Table~\ref{tab:skilfulness} records the mean skill scores for the GPs and DEs for each metric, averaged across the ten aircraft types. The GP is slightly more skilful than the DE when averaged across all aircraft types, although the DE performs better in generating the mean CAS and for for the more commonly occurring aircraft . Overall, the GP and DE have a 22.5\% and 20\% improvement over the baseline when averaged over all six metrics and ten aircraft types, split between around 25\% improvement in time and ROC predictions and 10\% for CAS.

\begin{figure}[htbp]
  \centering
  \begin{subfigure}{0.95\textwidth}
    \centering
    \includegraphics[width=\linewidth]{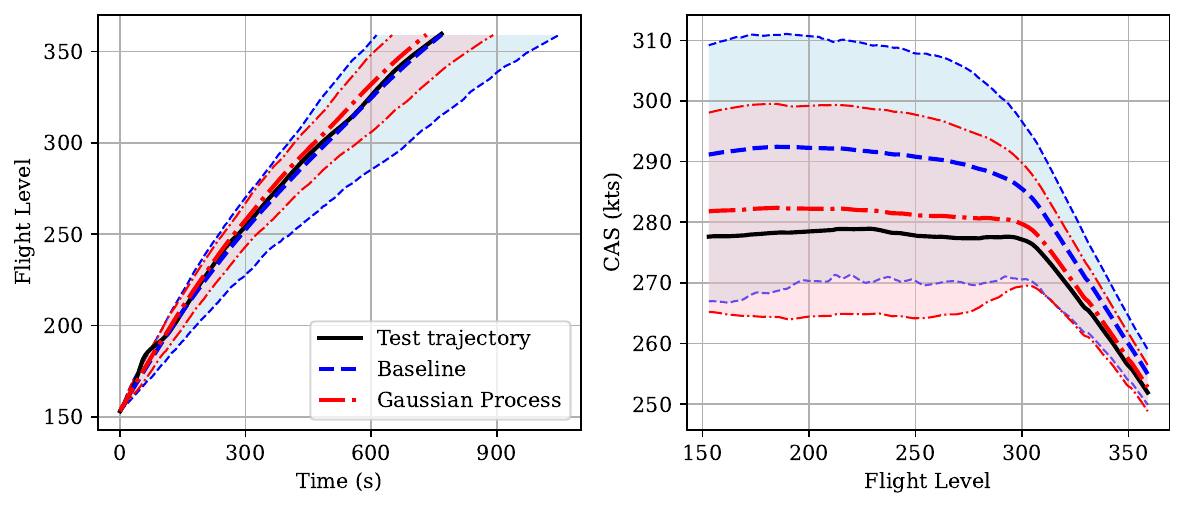}
    \caption{Test example 1}
    \label{fig:a320_traj_a}
  \end{subfigure}\hfill
  \begin{subfigure}{0.95\textwidth}
    \centering
    \includegraphics[width=\linewidth]{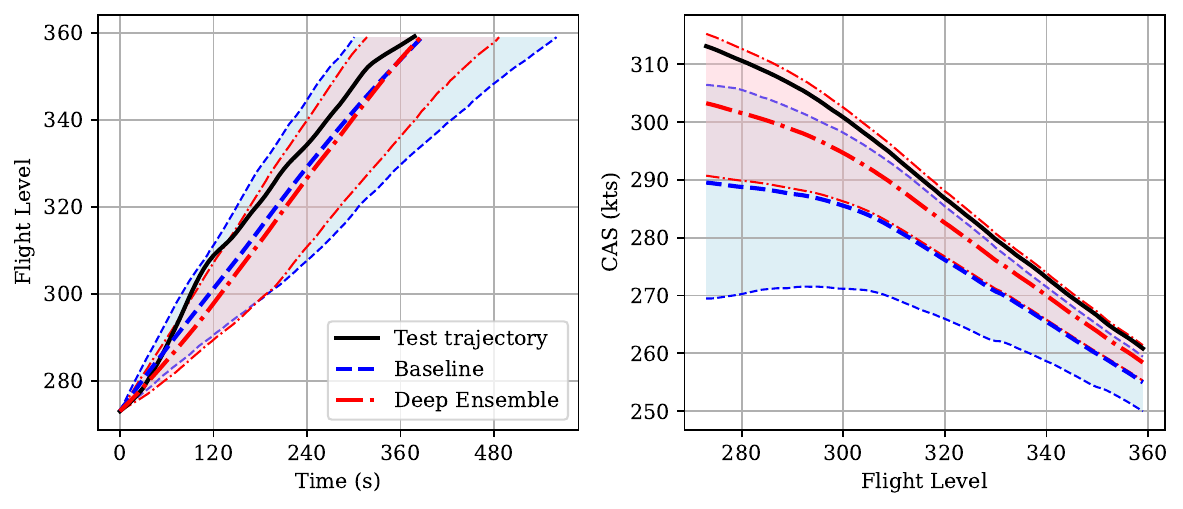}
    \caption{Test example 2}
    \label{fig:a320_traj_b}
  \end{subfigure}
  \caption{Test trajectory example, showing mean prediction from a conditioned model and baseline, and 95\% CIs.}
  \label{fig:a320_traj}
\end{figure}

All three of the investigated probabilistic models generate continuous distributions $p(\boldsymbol{y}|\boldsymbol{x})$. Sampling from this distribution can occasionally produce trajectories, when $\boldsymbol{y}$ is mapped to thrust and CAS functions and passed through BADA, that do not achieve the minimum legal climb rate of 500~ft/min in UK airspace. These are filtered out through rejection sampling. Figure~\ref{fig:rejection_rate} in Appendix A tracks the rate that samples are rejected across the aircraft types and probabilistic models, showing that the GP and DE have lower rejection rates than the baseline model. The average rejection rate across all aircraft for the GP is 2.8\%, the DE is 3.2\% and the baseline is 6.8\%. This is likely due to the GP and DE generating $p(\boldsymbol{y}|\boldsymbol{x})$ with lower variance compared to the Gaussian model. As a result less probability mass is distributed in regions of the fPCA weight space that lead to implausible trajectories. This can be clearly seen in Figure~\ref{fig:a320_traj}, which compares A320 sub-trajectories from the test dataset against the mean of the trajectories generated by the GP (blue) and baseline (red). The credible intervals (CIs) for the time at which every tenth flight level is reached is computed and shaded for the models. As is clear from Figure~\ref{fig:a320_traj_a}, the GP CIs are noticeably tighter than those of the baseline model, particularly for CAS. The effect is particularly pronounced in Figure~\ref{fig:a320_traj_a}, a longer trajectory that climbs through the CAS-Mach transition around flight level 300, compared to Figure~\ref{fig:a320_traj_b}, a shorter sub-trajectory above the CAS-Mach transition. 

\begin{table}[h!]
\centering
\caption{Skilfulness for each metric and variable, defined as $1 - \frac{S_\text{model}}{S_\text{baseline}}$, where $S$ is the mean score (RMSE or CRPS). Red cell color indicates negative skilfulness (ie not out performing baseline). The best model for each aircraft and metric is indicated by a box.}
\begin{tabular}{llcccccc}
\toprule
\toprule
Aircraft & Model & \multicolumn{3}{c}{RMSE skilfulness} & \multicolumn{3}{c}{CRPS skilfulness} \\
 &  & Time & Mean ROC & Mean CAS & Time & Mean ROC & Mean CAS  \\
\midrule

B738 & GP & 0.271 & 0.313 & 0.246 & 0.309 & 0.329 & 0.210 \\
 & DE &\ovalbox{0.324} & \ovalbox{0.367} & \ovalbox{0.269} & \ovalbox{0.348} & \ovalbox{0.377} & \ovalbox{0.238} \\ 

\midrule
 A320 & GP & 0.246 & 0.274 & 0.224 & 0.261 & 0.267 & 0.203  \\ 

 & DE & \ovalbox{0.248} & \ovalbox{0.297} & \ovalbox{0.245} & \ovalbox{0.289} & \ovalbox{0.286} & \ovalbox{0.218}  \\ 

\midrule
A319 & GP & 0.229 & 0.271 & 0.232 & 0.248 & 0.268 & 0.194   \\ 

 & DE & \ovalbox{0.233} & \ovalbox{0.305} & \ovalbox{0.255} & \ovalbox{0.285} & \ovalbox{0.299} & \ovalbox{0.216}   \\ 

\midrule
E190 & GP & \ovalbox{0.239} & \ovalbox{0.281} & \ovalbox{0.100} & \ovalbox{0.294} & \ovalbox{0.270} & 0.066 \\ 

 & DE & 0.230 & 0.275 & 0.094 & 0.283 & 0.251 & \ovalbox{0.068}  \\ 

\midrule
DH8D & GP & \ovalbox{0.391} & \ovalbox{0.369} & 0.200 & \ovalbox{0.396} & \ovalbox{0.359} & 0.187   \\ 

 & DE & 0.301 & 0.269 & \ovalbox{0.268} & 0.292 & 0.247 &  \ovalbox{0.241} \\ 

\midrule
A388 & GP & \ovalbox{0.324} & \ovalbox{0.273} & 0.129 & \ovalbox{0.307} & \ovalbox{0.222} & 0.065  \\ 

 & DE & 0.244 & 0.251 & \ovalbox{0.148} & 0.263 & 0.193 & \ovalbox{0.104}  \\ 

\midrule
E170 & GP & \ovalbox{0.041} & \ovalbox{0.098} & 0.040 & \ovalbox{0.131} & \ovalbox{0.086} & 0.012  \\ 

 & DE & \cellcolor[RGB]{255,179,179}-0.037 & 0.027 & \ovalbox{0.065} & 0.096 & \cellcolor[RGB]{255,179,179}-0.002 & \ovalbox{0.028}  \\ 

\midrule
PC12 & GP & \ovalbox{0.231} & \ovalbox{0.191} & \cellcolor[RGB]{255,179,179}\ovalbox{-0.047} & \ovalbox{0.236} & \ovalbox{0.190} & \cellcolor[RGB]{255,179,179}\ovalbox{-0.070} \\ 

 & DE & 0.170 & 0.103 & \cellcolor[RGB]{255,179,179}-0.092 & 0.155 & 0.073 & \cellcolor[RGB]{255,179,179}-0.133  \\ 

\midrule
C56X & GP & 0.157 & 0.222 & \cellcolor[RGB]{255,179,179}-0.022 & 0.181 & 0.216 & \cellcolor[RGB]{255,179,179}-0.079  \\ 

 & DE & \ovalbox{0.204} & \ovalbox{0.300} & \ovalbox{0.014} & \ovalbox{0.206} & \ovalbox{0.268} & \cellcolor[RGB]{255,179,179}\ovalbox{-0.017}  \\ 

\midrule
BE20 & GP & 0.249 & 0.205 & 0.175 & \ovalbox{0.262} & \ovalbox{0.274} & \ovalbox{0.111} \\ 

 & DE & \ovalbox{0.323} & \ovalbox{0.289} & \ovalbox{0.182} & 0.261 & 0.270 & 0.086  \\ 
 \midrule
 {Mean all} & GP & \ovalbox{0.238} & \ovalbox{0.250} & 0.128 & \ovalbox{0.262} & \ovalbox{0.248} & 0.090 \\ 
 types & DE & 0.224 & 0.248 & \ovalbox{0.145} & 0.248 & 0.226 & \ovalbox{0.105} \\
\bottomrule
\bottomrule
\end{tabular}

\label{tab:skilfulness}
\end{table}

\section{Discussion}

This paper has identified a set of 14 informative features for data-driven aircraft trajectory prediction that encode the operational context and local meteorological conditions encountered by an aircraft. Furthermore, it has been shown that conditioning a probabilistic trajectory generator on these features significantly improves the skilfulness of the model, both in terms of the accuracy of mean predictions and the calibration of the distribution of generated trajectories. Performance was quantified by a set of six metrics, with models conditioned on the informative features improving on a baseline probabilistic model by 20\% when averaged across the ten investigated aircraft types. Deep ensembles were compared against Gaussian processes for the task of generating trajectories within the proposed framework. Gaussian process models were generally more skilful for infrequently occurring aircraft types. However, for more frequently common aircraft types such as the B738 and A320, with more than 20,000 trajectories in the training dataset, deep ensembles gave best performance. This paper focuses on probabilistic models of climbing aircraft within en route airspace. Future work could analyse different phases of flight, as well as extending the study to terminal airspace, which is more procedural and may require additional features to describe the operational context. 

\FloatBarrier

\section*{Appendix}
\subsection{Reduced order basis implementation details}\label{app:fpca}
Section~\ref{sec:low_order} outlines the fPCA basis which was used as a low-dimensional space on which to machine learn functional enhancements to BADA. In contrast to previous work that has applied fPCA to whole aircraft trajectories (see, e.g. \cite{pepper2023learning, hodgkin2025probabilistic, nicol2013functional, pepper4984556probabilistic}), in this paper we wish to project individual portions of a climbing trajectory to an fPCA basis, which requires additional processing. Given an aircraft trajectory comprising $n$ radar blips, $\mathcal{T} = [\big(h^{(0)}, \frac{dh^{(0)}}{dt}, \hat{V}_{CAS}^{(0)}\big), \dots, \big(h^{(n)},\frac{dh^{(n)}}{dt}, \hat{V}_{CAS}^{(n)}\big)]$, that spans the grid of flight levels $\boldsymbol{g}$, the following steps are followed: 

\begin{enumerate}
    \item Remove radar blips with $\frac{dh}{dt}<500\text{ft/min}$, the legal minimum climb rate.
    
    \item Equation $\eqref{eq:bada_rocd}$ may be rearranged to give an expression for thrust as a function of $h$. Use this expression to compute an inferred drag for the remaining radar blips, yielding the augmented trajectory $\hat{\mathcal{T}} = [\big(h^{(0)}, \frac{dh^{(0)}}{dt}, V_{CAS}^{(0)}, \hat{T}_{HR}^{(0)}\big), \dots, \big(h^{(n)},\frac{dh^{(n)}}{dt}, V_{CAS}^{(n)}, \hat{T}_{HR}^{(n)}\big)]$.
    
    \item The dataset of augmented trajectory data containing $n_t$ trajectories, denoted $\mathcal{D}=[\hat{\mathcal{T}}_1,\dots, \hat{\mathcal{T}}_{n_t}]$ can then be used to determine the basis functions for $\hat{T}_{HR}$ and $\hat{V}_{CAS}$ through fPCA. 

    \item Sub-divide the trajectories into sub-trajectories in which the time difference between consecutive radar blips matches the refresh rate of the radar. This dataset is denoted $\mathcal{D}_{s} = [\mathcal{S}_1, \dots, \mathcal{S}_{n_s}]$.
    
\end{enumerate}

Having computed a dataset of sub-trajectories and basis functions, optimal model weights $\boldsymbol{\alpha}$ and $\boldsymbol{\beta}$ are found for each $\mathcal{S}_i, \;i\in[1,n_s]$ through an optimisation process: 
\begin{align}
\boldsymbol{\alpha}_i^* = \arg\min_{\boldsymbol{\alpha}_{1:i}} 
\left\| \left(\mathbf{z}_{\text{obs}} - \mu_{T_{HR}}(\boldsymbol{x}) -\sum_{j=1}^{i-1}\boldsymbol{\alpha}_j \phi_j(\boldsymbol{x})\right)
\right\|^2 + \lambda \boldsymbol{\alpha}_i^2,
\label{eq:opt1}
\end{align}
with a similar optimisation process defined for the weights $\boldsymbol{\beta}$ in \eqref{eq:fpca_cas}. The second term on the right hand side is an optional regularisation term, although for the preliminary results described below it is set as $\lambda =0$. This yields the set of target vectors $\mathcal{D}_{y} = [\boldsymbol{y}_1,\dots, \boldsymbol{y}_{n_s}]$, where $\boldsymbol{y}_i=[\boldsymbol{\alpha}^*_i, \boldsymbol{\beta}^*_i]^\top$.

\begin{figure}
    \centering
    \includegraphics[width=0.4\linewidth]{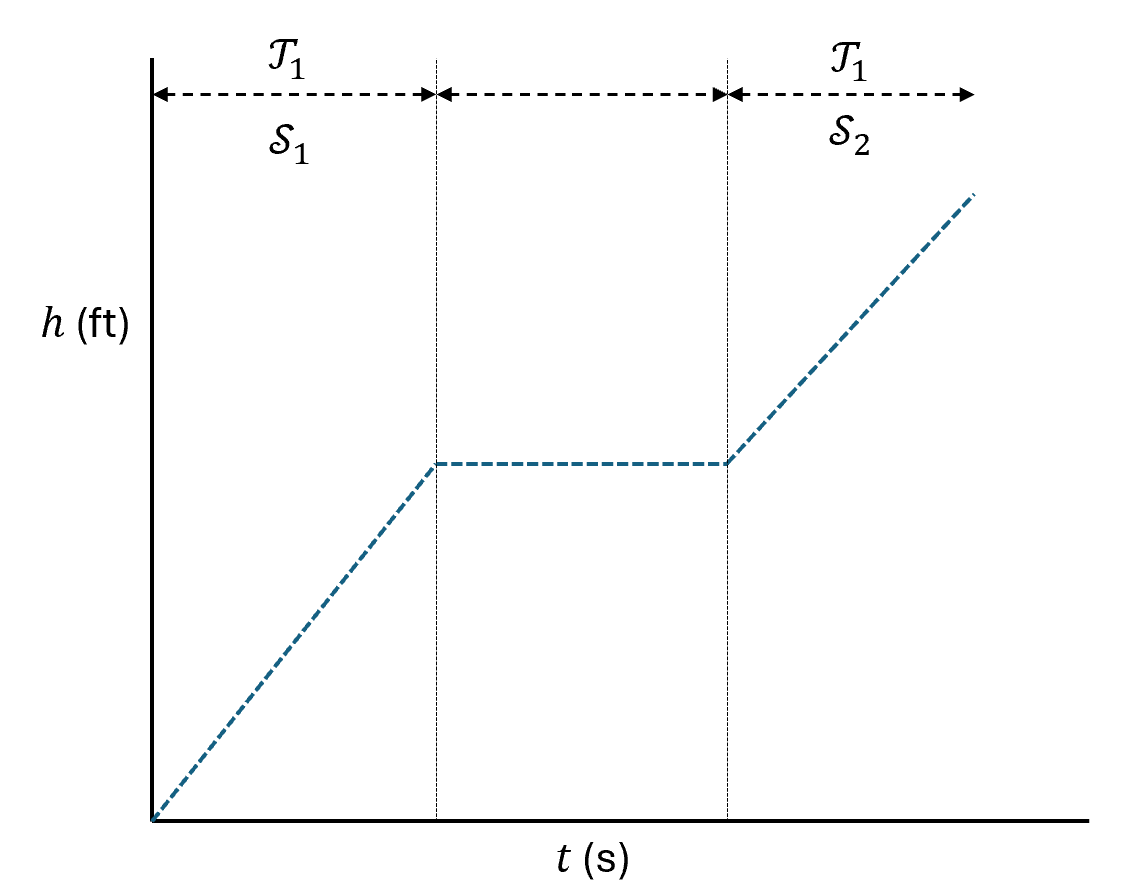}
    \caption{Schematic illustrating the difference between trajectories, $\mathcal{T}_i$, and sub-trajectories $\mathcal{S}_j$, as defined in this paper. }
    \label{fig:notation}
\end{figure}

\subsubsection{Explained Variance}\label{app:b738_coeffs}
\begin{figure}[ht]
    \centering
  \includegraphics[width=0.6\linewidth]{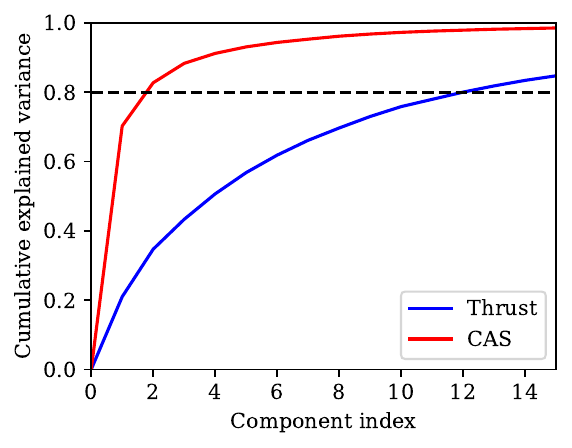}
    \caption{Cumulative explained variance for the fPCA projection for the B738 thrust and CAS functions, against number of coefficients in the expansion.}
   \label{fig:expvar}
\end{figure}

The fPCA expansions in \eqref{eq:fpca_sum} were truncated to $n_\alpha$ and $n_\beta$ basis functions. Individual fPCA basis functions were developed for each aircraft type. As can be seen from Table~\ref{tab:features} and Figure~\ref{fig:expvar}, which illustrates the cumulative explained variance against component number for the B738, the thrust expansion generally required more terms than the CAS function to achieve 80\% explained variance. 

\FloatBarrier
\subsection{Rejection rates across probabilistic models}\label{sec:test_appendix}
\begin{figure}[h]
    \centering
    \includegraphics[width=0.85\linewidth]{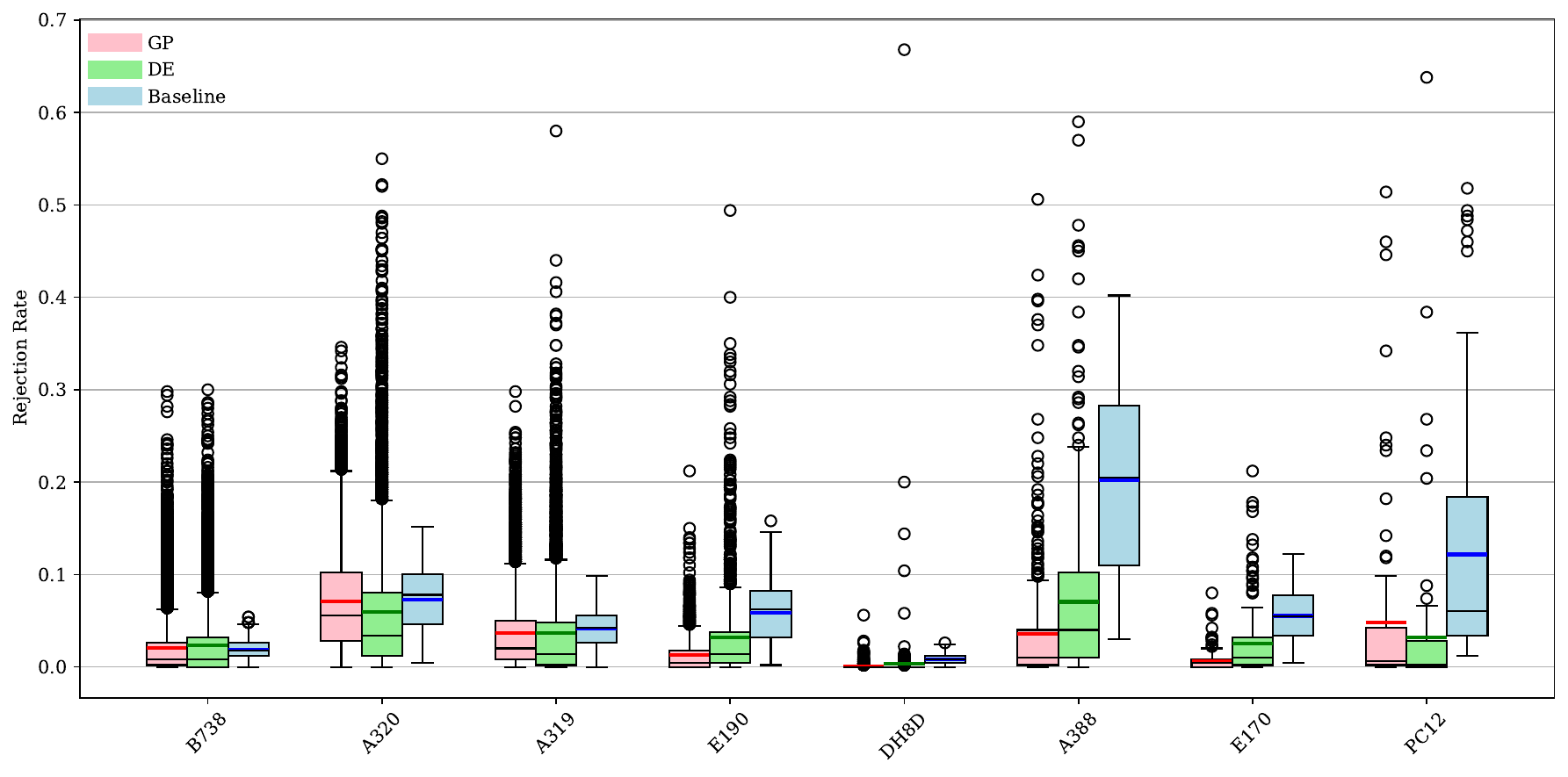}
    \caption{Rejection rates for each aircraft.}
    \label{fig:rejection_rate}
\end{figure}

Figure~\ref{fig:rejection_rate} is a set of boxplots comparing the rate at which implausible trajectories were generated across models and aircraft types. It is clear from the figure that the GP and DE has significantly lower rejection rates, particularly for the A388. 

\FloatBarrier
\newpage
\section*{Acknowledgments}
The work described in this paper is primarily funded by the grant “EP/V056522/1: Advancing Probabilistic Machine Learning to Deliver Safer, More Efficient and Predictable Air Traffic Control” (aka Project Bluebird), an EPSRC Prosperity Partnership between NATS, The Alan Turing Institute, and the University of Exeter. 

\bibliography{ref}

\end{document}